\documentclass[12pt]{article}
\usepackage[margin=1.8cm]{geometry}                
\usepackage{graphicx}
\usepackage{amssymb}
\usepackage{amsmath}
\usepackage{amsthm}
\usepackage{MnSymbol}
\usepackage{hyperref}
\usepackage{epstopdf}
\usepackage{tikz}
\usepackage{comment}
\usepackage{cite}
\usepackage{appendix}
\usepackage{lipsum}
\usepackage{psfrag}
\usepackage{pstool}
\usepackage{caption}

\def\e{{\rm e}}
\def\ii{{\rm i}}

\def\be{\begin{equation}}
\def\ee{\end{equation}}
\def\bea{\begin{eqnarray}}
\def\eea{\end{eqnarray}}

\def\ket#1{|#1\rangle}
\DeclareMathOperator{\sign}{sign}

\usepackage[font=small, labelfont=bf, width=0.8\textwidth]{caption}

\title{Integrable supersymmetric chain without particle conservation}
\author{Jan de Gier$^1$,  Gyorgy Z. Feher$^{1,2}$ Bernard Nienhuis$^2$, \\
 and Magdalena Rusaczonek$^2$\bigskip \\
\parbox{0.6\textwidth}{\center
\small $^1$\textit{ARC Centre of Excellence for Mathematical and Statistical Frontiers, School of Mathematics and Statistics, The University of Melbourne, VIC 3010, Australia}
\medskip\\
$^2$\textit{Instituut voor Theoretische Fysica, University of Amsterdam,
  P.O.Box 94485, 1090 GL Amsterdam, The Netherlands}}\bigskip}
\small\date{\small\today}                                          

\begin{document}
\maketitle
\abstract{We introduce a new integrable supersymmetric lattice chain which  violates fermion conservation and exhibits fermion-hole symmetry. The model displays exponential degeneracy in every eigenstate including the groundstate. This degeneracy is expressed in the possibility to create any number of zero modes reminiscent of Cooper pairs. }
\let\thefootnote\relax\footnotetext{Email: \texttt{jdgier@unimelb.edu.au,\, B.Nienhuis@uva.nl,\, G.Feher@uva.nl}}

\section{Introduction}

Supersymmetric lattice models have been studied starting from the $\mathcal{N}=1$ supersymmetry in the tricritical Ising model \cite{FriadenQiuShenker, KastorMartinecShenker, Qiu} and the fully frustrated XY-model \cite{Foda}. In the introduction of \cite{SaleurWarner} the interested reader can find further references to early supersymmetric lattice models.  

Our work goes back to\cite{FendleySchoutensBoer}, where a certain one-dimensional fermionic lattice model was constructed based on two supercharges (generators of supersymmetry) $Q$ and its hermitian conjugate $Q^\dagger$.  An alternative more general approach is to work with $Q_1 = Q+Q^\dagger$ and $Q_2 =\ii( Q-Q^\dagger )$ which naturally admits an extension of the number of generators, $\mathcal{N}$, which in this case is $\mathcal{N}=2$. 
Supersymmetry is based on the property that $Q^2=0$ (in the general approach $Q_j^2 = Q_k^2$), which has many consequences for the spectrum of the Hamiltonian which is defined as $H = \{Q,Q^\dagger\}$.  In particular there are no negative energy states, and all positive energy states come in doublets with the same eigenvalue, but differing in number of fermions by one.
States with zero energy, the lowest possible, can be highly degenerate, but form singlets with respect to supersymmetry.
By the specific choice of $Q$, the model in \cite{FendleySchoutensBoer} has  a repulsive hardcore potential between the fermions, i.e. two neighbouring particles have to be separated by one empty site. In \cite{FendleyNienhuisSchoutens} this work has been continued by generalizing the interaction, i.e. not a single, but $k$-long strings of fermions have to be separated by an empty site.

We modify the supersymmetric model of
\cite{FendleySchoutensBoer} in a different way: we restore the
particle--hole symmetry, by symmetrizing the building elements of the model, $Q$ and $Q^{\dagger}$.  In the original definition
$Q$ creates a solitary fermion: a fermion is created only on sites of which the neighbors are empty.  We simply symmetrize this action with respect to the particle--hole symmetry.  As an additional term in $Q$, we introduce the  
operator that creates a ``solitary hole'', i.e. annihilate a fermion between 
 neighbouring sites that are occupied. 
Because in the original model the particle--hole symmetry is strongly broken by the hardcore repulsion between fermions, 
this modification changes the nature of the model to a large
extent.   For example, the fermion number is no longer conserved.
Instead, we identify domain walls with Majorana--like properties, as
conserved objects.  
But most surprisingly, this modification of $Q$ does not violate  supersymmetry, i.e. the property that $Q^2=0$.
On the contrary, the high degeneracy of the zero energy state, a common
feature of supersymmetric models \cite{HS10,MSP03,FS05,E05}, is no longer limited to the ground
state, but in
this model all the eigenvalues exhibit an {\em extensive} degeneracy.  While in the original model all energy levels are two-fold degenerate, in the model investigated here, the degeneracy is a power of two with an exponent growing linearly in the system size.
This property suggests a much higher symmetry and, because this symmetry
is not at all evident, it was our prime motivation to study the model.
We investigate the reason for the large degeneracy, and provide an answer in terms of symmetries and zero energy Cooper-pair like excitations as in \cite{FS}. We further show that the system's energy gap scales as $\sim 1/L^2$ which is usually associated to classical diffusive modes.

The paper is organized as follows. In Section \ref{sec:ModelDef} we define the model, and introduce the most important operators and notation. The model turns out to be solvable by nested Bethe Ansatz \cite{Baxter70}. In Section~\ref{sec:Symmetries} we expose the Bethe equations without derivation, the large degeneracy of the model, and  the associated symmetry operators. In Section \ref{sec:Consequences} we provide some detailed examples of consequences of the symmetry operators, and in Section \ref{sec:BetheAnsatz}  we derive the Bethe Ansatz equations for the model. Our approach is educational: we gradually look at more complicated cases, and derive the Bethe-equations for them. 

\subsection{Model definition}
\label{sec:ModelDef}
In this section we define our model. We introduce the usual fermionic operators, fermionic Fock-space on lattice, and the supersymmetric generators $Q$ and $Q^{\dagger}$. The Hamiltonian is defined in terms of these generators. As we will see, the fermion number is not a conserved quantity, so we introduce an other notation, based on different conserved quantities (domain walls), useful for solving the model with Bethe ansatz.

We define the usual fermionic creation and annihilation operators $c_j$ and $c_j^\dagger$, acting on site $j$, satisfying the anti-commutation relations
\be
\{c_i^\dagger,c_j\}=\delta_{ij},\qquad \{c_i^\dagger,c_j^\dagger\}=\{c_i,c_j\}=0.
\ee
The on site fermion-number and hole-number operators are defined as
\be
n_i = c_i^\dagger c_i, \qquad p_i=1-n_i.
\ee
The number operator, $\mathcal{N}_i$, of fermions on positions $1$ to $i$ and the total fermion number operator, $\mathcal{N}_{\rm F}$, are defined as 
\be
\mathcal{N}_i = \sum_{j=1}^i n_j, \qquad \mathcal{N}_{\rm F}:=\mathcal{N}_L.
\ee
These operators act in a fermionic Fock space spanned by ket vectors of the form
\be
\ket{\mathbf{\tau}} = \prod_{i=1}^L \left(c_i^\dagger\right)^{\tau_i} \ket{\mathbf{0}},
\ee 
where the product is ordered and $\ket{\mathbf{0}}$ is the vacuum state defined by $c_i  \ket{\mathbf{0}} =0$ for $i=1,\ldots, L$.  The label $\mathbf{\tau}=\{\tau_1,\ldots,\tau_L\}$, with $\tau_i=1$ if there is a fermion at site $i$ and $\tau_i=0$ if there is a hole. Hence we have
\be
n_i \ket{\mathbf{\tau}}= \tau_i \ket{\mathbf{\tau}}.
\ee

We consider a one-dimensional supersymmetric lattice model analogous to 
\cite{FendleySchoutensBoer}, but satisfying  fermion-hole symmetry. For that purpose define the operators $d_i^\dagger$ and $e_i$ by
\be
d_i^\dagger = p_{i-1}c_i^\dagger p_{i+1},\qquad e_i =  n_{i-1}c_i n_{i+1}. 
\ee
Hence $d^\dagger_i$ creates a fermion at position $i$ provided all three of positions $i-1$, $i$ and $i+1$ are empty. Similarly, $e_i$ annihilates a fermion at position $i$ provide $i$ and its neighbouring sites are occupied, i.e.
\begin{align}
d^\dagger_i \ket{\tau_1\ldots \tau_{i-2}\, 000\,\tau_{i-2}+\ldots\tau_L} &= (-1)^{\mathcal{N}_{i-1}}\ket{\tau_1\ldots\tau_{i-2}\, 010\,\tau_{i+2}\ldots\tau_L},\\
e_i \ket{\tau_1\ldots\tau_{i-2}\, 111\,\tau_{i+2}\ldots\tau_L} &= (-1)^{\mathcal{N}_{i-1}}\ket{\tau_1\ldots\tau_{i-2}\, 101\,\tau_{i+2}\ldots\tau_L},
\end{align}
while these operators kill all other states.

We now define a supersymmetric Hamiltonian $H$ for fermions on a chain, by
\be
H=\{Q^\dagger,Q\},\qquad Q=\sum_{i=1}^L (d_i^\dagger + e_i),\qquad Q^2=0. 
\ee
This is a simple variation of the supercharge considered in 
\cite{FendleySchoutensBoer},   $Q=\sum_{i=1}^L d_i^{\dagger}$, by adding to it 
the fermion--hole symmetric partner of $d_i^{\dagger}$ thus restoring that symmetry.
It is unexpected that this variation respects the requirement $Q^2=0$ of supersymmetry.

The Hamiltonian splits up naturally as a sum of three terms, the first term consists solely of $d$-type operators, the second solely of $e$-type operators and the third contains mixed terms.
\be
H=H_I +H_{II}+H_{III},
\label{eq:modeldef}
\ee
\begin{align}
H_I &= \sum_i \left( d_i^\dagger d_i + d_i d_i^\dagger + d_{i}^\dagger d_{i+1} + d_{i+1}^\dagger d_i\right),\nonumber\\
H_{II} &= \sum_i \left( e_i e_i^\dagger + e_i^\dagger e_i + e_i e_{i+1}^\dagger + e_{i+1} e_{i}^\dagger \right),\\
H_{III} &= \sum_i \left( e_i^\dagger d_{i+1}^\dagger\ + d_{i+1} e_i + e_{i+1}^\dagger d_i^\dagger + d_i e_{i+1}\right), \nonumber
\end{align}
where we use periodic boundary conditions
$
c_{i+L}^\dagger = c_{i}^\dagger.
$
Because the $d$ and $e$ are not simple fermion operators, they do not satisfy the canonical anticommutation relations.  As a result this bilinear Hamiltonian can not be diagonalized by taking linear combinations of $d$, $e$, $d^\dagger$ and $e^\dagger$.

The Hamiltonian $H_I$ was considered in \cite{FendleySchoutensBoer} and is obtained when operators $e_i^\dagger$ and $e_i$ are not included in $Q$. In this case the model is equivalent to the integrable spin-1/2 quantum XXZ spin chain with $\Delta=-1/2$ and with variable length. The groundstate of this model exhibits interesting combinatorial properties.

The additon of the operator $e_i$ adds an obvious `fermion-hole' symmetry $d_i^\dagger \leftrightarrow e_i$ to the model which was our original motivation. As we will see, this symmetry results in a surprisingly large degeneracy across the full spectrum of $H$. Moreover, the new model \eqref{eq:modeldef} unexpectedly turns out to be integrable, as we will show below.

Note that the Hamiltonians $H_I$ and $H_{II}$ each contain only number operators and hopping terms and thus conserve the total number of fermions. The third Hamiltonian $H_{III}$ breaks this conservation law. For example, the term $e_i^\dagger d_{i+1}^\dagger$ sends the state $\ket{\ldots 1000\ldots}$ to  $\ket{\ldots 1110\ldots}$, thus creating two  fermions. Hence the fermion number is not conserved and not a good quantum number. However, the number of interfaces or domain walls between fermions and holes is conserved, and we shall therefore describe our states in terms of these.

%

\subsection{Domain walls}
\label{se:domainwalls}
We call an interface between a string of 0's followed by a string of 1's a 01-domain wall, and a string of 1's followed by a string of 0's, a 10-domain wall. For example, the following configuration contains six domain walls (we consider periodic boundary conditions), three of each type and starting with a 01-domain wall,
\[
000\Big| 11\Big| 000\Big| 1\Big| 0000\Big| 111\Big|
\]

Let us consider the effect of the various terms appearing in \eqref{eq:modeldef}. As already discussed in an example above, the terms in $H_{III}$ correspond to hopping of domain walls and map between the following states
\be
\ket{\ldots 1\Big|000\ldots} \leftrightarrow  \ket{\ldots 111\Big|0\ldots},\qquad  \ket{\ldots 0\Big|111\ldots} \leftrightarrow  -\ket{\ldots 000\Big|1\ldots},
\label{eq:process1}
\ee
where the minus sign in the second case arises because of the fermionic nature of the model. Hopping of a domain wall always takes place in steps of two, so parity of position is conserved. Aside from their diagonal terms, $H_I$ and $H_{II}$ correspond to hopping of single fermions or holes, and therefore to hopping of \textit{pairs} of domain walls. They give rise to transitions between the states
\be
\ket{\ldots 0\Big|1\Big|00\ldots} \leftrightarrow  \ket{\ldots 00\Big|1\Big|0\ldots},\qquad  \ket{\ldots 1\Big|0\Big|11\ldots} \leftrightarrow  -\ket{\ldots 11\Big|0\Big|1\ldots},
\label{eq:oddprocess}
\ee
Note that in these processes the total parity of positions of interfaces is again conserved, i.e. all processes in $H$ conserve the number of domain walls at even and odd positions separately.

Finally, the diagonal term $\sum_i (d_i^\dagger d_i + d_i d_i^\dagger + e_i^\dagger e_i + e_i e_i^\dagger)$ in $H_{I}$ and $H_{II}$ counts the number of  $010$, $000$, $111$ and $101$ configurations. In other words it counts the number of pairs of second neighbour sites that are both empty or both occupied,
\be
\sum_i (d_i^\dagger d_i + d_i d_i^\dagger + e_i^\dagger e_i + e_i e_i^\dagger) = \\\sum_i (p_{i-1}p_{i+1} + n_{i-1}n_{i+1}).
\ee
This is equivalent to counting the total number of sites minus twice the number of domain walls that do not separate a single fermion or hole, i.e. twice the number of well separated domain walls. 

Since the number of odd and even domain walls is conserved, the Hilbert space naturally breaks up into sectors labelled by $(m, k)$, where $m$ is the total number of domain walls, and $k$ the number of odd domain walls. 

\section{Symmetries}
\label {sec:Symmetries}
The most remarkable feature of the model introduced in Section~\ref{sec:ModelDef} is the high degeneracy not only of the ground state, but of all the eigenvalues of the Hamiltonian.  The number of different eigenvalues and the typical degeneracy both grow exponentially with the system size.  Aside from some staggering with the system size modulo 4, the growth rate of the degeneracy and of the number of levels appears similar.

In this section we show that the model  possesses symmetries which explain the large degeneracy of the energy levels. Fermions and holes are treated on the same footing and consequently the model is symmetric under the exchange of fermions and holes. Even though the number of fermions is not conserved, the fermion number can only change by two, so the parity of the number of fermions is conserved. The model is also invariant under the exchange of domain wall with non-domain walls.  This symmetry interchanges the off-diagonal terms of $H_I$ and $H_{II}$ with $H_{III}$.   Below we will describe further symmetries, first those that we can describe by simple real-space operators.

In addition to these, the model possesses a symmetry in momentum space due to the possibility of creating and removing zero mode Cooper pairs. This symmetry leads to an extensive degeneration of the ground state and other eigenstates. 

As an indication of the high degeneracy, we list  for  system size $L$ up to 12,  the number of groundstates $G$, and the number of different energy levels $\ell$, see Table~\ref{tab:degeneracy}. As the model respects particle hole symmetry, it makes sense to consider besides periodic also antiperiodic boundary conditions, defined by $c_j=c^{\dagger}_{L+j}$. We give the results for this boundary condition as well,  because the two lists together give a better idea of the growth of these numbers.

While the mean degeneracy can be seen from the number of energy levels, we remark that almost all degeneracies that we see are powers of two.  All this seems to indicate a high symmetry, which this paper aims to explain.

\begin{table}[h]
\begin{center}
\begin{tabular}{r|rr|rr}
    & \multicolumn{2}{c|}{periodic}& \multicolumn{2}{c}{antiperiodic}\\
    \hline
$L$ &  $G$ & $\ell$&   $G$ & $\ell$\\ \hline
4   & 8    & 2   &  4  & 3   \\
5   & 8    & 6   &  8  & 5   \\
6   & 0    & 4   &  16 & 4   \\
7   & 16   & 15  &  16 & 14  \\
8   & 32   & 7   &  16 & 20  \\
9   & 32   & 54  &  32 & 54  \\
10  & 0    & 46  &  64 & 94  \\
11  & 64   & 204 &  64 & 210 \\
12  & 128  & 80  &  64 & 201 
\end{tabular}
\caption{The degeneracy $G$ of the groundstate, the number of energy levels $\ell$,
  for periodic and antiperiodic boundary conditions}
\label{tab:degeneracy}
\end{center}
\end{table}

\subsubsection*{Supersymmetry}
Obviously, by construction the supersymmetry generators commute with the Hamiltonian,
\begin{align}
 [ H,\, Q ]= 0, \qquad  [ H,\, Q^\dagger ] = 0.
\end{align}
The supercharges $Q$ and $Q^\dagger$ are operators that add or remove a fermion, which means that they add or remove two neighbouring domain walls, one even and one odd, respectively, i.e.
\be
Q:\ (m,k)\mapsto (m+2,k+1),\qquad Q^\dag:\ (m,k)\mapsto (m-2,k-1),
\ee
where $(m,k)$ denotes the sector with $m$ domain walls of which $k$ are odd.

\subsubsection*{Domain wall number conservation and translational symmetry}
Two obvious symmetries are the total number of domain walls and translational symmetry due to the periodic boundary conditions for even system sizes. The domain wall number operator $\mathcal{D}$ commutes with $H$, $[H,\mathcal{D}]=0$ and so does the translation operator $T$.

\subsubsection*{Particle parity symmetry}
The total number of fermions is not conserved as both $H_{III}$ changes the fermion number.  We denote the fermion parity operator by $P$, which acts on pure states $\ket{\tau_1,\ldots,\tau_L}$ as
\be
P \ket{\tau_1,\ldots,\tau_L} = (-1)^{\mathcal{N}_L} \ket{\tau_1,\ldots,\tau_L} .
\ee  
Since $Q$ and $Q^\dag$ change parity  the supersymmetry generators anti-commute with $P$,
\be
\{Q,P\}=\{Q^\dag,P\}=0,
\ee
from which it is simple to show that $[H,P]=0$.

\subsubsection*{Particle -- hole symmetry} 

Introduce the operator

\begin{equation}
 \Gamma= \prod_{i=1}^L \gamma_i, \qquad \gamma_i=c_i + c_i^\dagger, 
\end{equation}
in terms of the Majorana fermions $\gamma_i$. This operator acts on a fermionic state $\ket{\tau}$ by exchanging the holes and fermions, and it is easy to see, that this is a symmetry of the model:

\begin{equation}
 [ H,\, \Gamma ] = 0.
\end{equation}
In fact one can show that $\Gamma$ either commutes or anti-commutes with the supersymmetry generators
\be
Q\Gamma+(-1)^L \Gamma Q=0,\qquad Q^\dag \Gamma +(-1)^L \Gamma Q^\dag=0. 
\ee 

\subsubsection*{Domain wall -- non-domain wall symmetry}
For even system sizes it is not hard to see that we can expect a domain wall (DW) -- non domain wall (nonDW) symmetry. The processes described in \eqref{eq:process1} and \eqref{eq:oddprocess} are interpreted as movement of a single DW or a bound double DW, but equivalently they can be interpreted as the movement of a bound double nonDW, and single nonDW respectively. The DW-nonDW exchange operator can be written as
\begin{equation}
E = \prod_{i=1}^{L/2} (c_{2i}^{\vphantom{\dagger}} - c_{2i}^\dagger),
\end{equation}
which satisfies the commutation relations
\begin{align}
EQ &= Q^\dagger E, \qquad EQ^\dagger = QE, \qquad EH = HE.
\end{align}
The DW -- nonDW symmetry interchanges the sectors $(m,k)$ with $(L-m,L/2-m+k)$.

\subsubsection*{Shift symmetry}
\label{sec:Shift}
There is a further symmetry, defined by the operator $S$:
\begin{equation}
 S=\sum_{i=1}^L n_{i-1} \gamma_i p_{i+1} + p_{i-1} \gamma_i n_{i+1},\qquad \gamma_i=c_i + c_i^\dagger.
\end{equation}
The operator $S$ shifts one of the domain walls either to the left or to the right by one. It is easy to see from the definition, that $S$ is self-adjoint, in fact, each summand is self-adjoint. By explicit computation, we can show that $S$ anticommutes with $Q$ and $Q^\dagger$,
\begin{equation}
\{Q,\,S\}=0,\quad \{Q^\dagger,\,S\}=0,\qquad [H,S]=0.
\end{equation}

This defines a symmetry of the model which relates the sector $(m,k)$ with the sectors $(m,k\pm 1)$. 

\subsubsection*{Reflection symmetry of the spectrum for $L=4n$}

It is easy to prove, that for $L=4n, \, n \in \mathbb{N}$, the groundstate energy is $\Lambda_{0}=0$ and the highest energy level is given by $\Lambda_{\text{max}}=L$.  We have observed, that the spectrum has a reflection symmetry, i.e., if there is an energy level with energy $\Lambda=L-\Delta \Lambda$, then there is one with $\widetilde{\Lambda}= \Delta \Lambda$. The degeneracy for these two mirrored levels is the same. These two energy levels are related by an operator defined in the following way. Let
\be
\delta_j =\ii\, (c_j-c_j^\dag),\qquad \delta_j^\dag = \delta_j.
\ee
Then define
\begin{equation}
 M = \prod_{i=0}^{n-1}  \delta_{4i+1} \delta_{4i+2}  = (-1)^{n} \prod_{i=0}^{n-1} (c_{4i+1} - c_{4i+1}^\dagger ) (c_{4i+2} - c_{4i+2}^\dagger ).
\end{equation}

The operator $M$ is (anti)hermitian depending on the parity of $n$, and squares to a multiple of the identity,
\begin{equation}
 M^{\dagger} = (-1)^n M, \qquad M^2 = (-1)^n \mathbb{I}.
\end{equation}
The mirroring property is encoded in $M$ in the following way,
\begin{equation}
 M (L  \mathbb{I} - H)  =  H M,
\end{equation}
which means that for every eigenvector there is mirrored pair,
\begin{equation}
H\ket{\Psi} = \Lambda \ket{\Psi}  \Leftrightarrow H M \ket{\Psi} = (L-\Lambda)M \ket{\Psi}.
\end{equation}
A good example of this pairing is to take the pseudo-vacuum $\ket{000 \ldots 0}$. This state maps into a half filled true ground state, i.e. into $\pm \ket{110011001 \ldots 100}$ (where the sign depends on $n$). 

\subsubsection*{Antiperiodic boundary conditions and reflection symmetry of the spectrum for $L=4n-2$}

The reflection symmetry can be extended to antiperiodic boundary conditions, and for $L=4n-2$ systems, we can relate the antiperiodic spectrum with the periodic one by the mirroring. 
Introduce antiperiodic boundary conditions, which we will use only in this section:
\begin{equation}
 c_{i+L}^\dagger = c_i.
\end{equation}
This modifies the Hamiltonian, which we will denote by $H_{ap}$. The antiperiodic Hamiltonian's spectrum has the same reflection symmetry as the periodic for $L=4n$. The definition of $M$ is independent of the boundary condition, so we can write 
\begin{equation}
M (L  \mathbb{I} - H^{(L=4n)}_{ap})  =  H^{(L=4n)}_{ap} M,
\end{equation}
where for clarity we emphasized the system size $L=4n$.

We have observed, that for $L=4n-2$, the periodic and the antiperiodic spectrum is related by the previous reflection, i.e. if there is a state of $H_{ap}$ with energy $\Lambda_{ap}$, there is a corresponding state of $H$ with energy $L-\Lambda_{ap}$. The largest energy for $H$ is $\Lambda_{p,max} = L$, corresponding to the antiperiodic GS with $\Lambda_{ap,GS} = 0$, which reflection is realized by the next operator equation:
\begin{equation}
M (L  \mathbb{I} - H^{(L=4n-2)}_{ap})  =  H_{p}^{(L=4n-2)} M,
\end{equation}
where we stressed the periodic Hamiltonian by $H_p$.

The last relation is easy to understand intuitively: For $L=4n-2$, $H_p$ has the largest eigenvalue equal to $L$ corresponding to e.g. the state $\ket{000 \ldots 00}$. This is mapped to $\ket{1100110...0011}$, where the first and the last two sites are all occupied. But since the boundary conditions are antiperiodic, this GS is analoguous to the periodic GS $\ket{0011..0011}$ for $L=4n$.

\subsubsection*{Zero mode Cooper pairs}

The Hamiltonian $H$ in \eqref{eq:modeldef} is diagonalisable using Bethe's ansatz. We derive the Bethe equations and present the explicit form of the Bethe vectors in Section~\ref{sec:BetheAnsatz}.  Here we present the Bethe equations to elucidate a large symmetry which is most obvious in momentum space. 

Note that there are two type of pseudo-particles, namely domain walls and odd domain walls. To diagonalise \eqref{eq:modeldef} we therefore employ a nested Bethe ansatz. Each domain wall is associated with a Bethe-root $z_j$, where $\log z_j$ is proportional to the momentum of the domain wall, and each odd domain wall is associated with an additional, nested Bethe-root $u_l$. In Section~\ref{sec:BetheAnsatz} we show that in the sector $(m,k)$, and for even system sizes $L$, the eigenvalue of $H$ is given by
\be
\Lambda =L+ \sum_{i=1}^{m} (z_i^2+z_i^{-2}-2).
\label{eq:eigval}
\ee
where the set of $z_1, z_2, \ldots, z_m$ and $u_1, \ldots, u_k$ satisfies the equations,
\begin{align}
z_j^L & =\pm \ii^{-L/2} \prod_{l=1}^k \frac{u_l-(z_j-1/z_j)^2}{u_l+(z_j-1/z_j)^2},\qquad j=1,\ldots,m
\label{eq:bae1}\\
1 &=  \prod_{j=1}^{m} \frac{u_l-(z_j-1/z_j)^2}{u_l+(z_j-1/z_j)^2},\qquad l=1,\ldots,k,
\label{eq:bae2}
\end{align}
where the $\pm$ is the same for all $j$.

\begin{figure}[t]
\centering
\begin{minipage}{.5\textwidth}
  \centering
  \includegraphics[width=.7\linewidth]{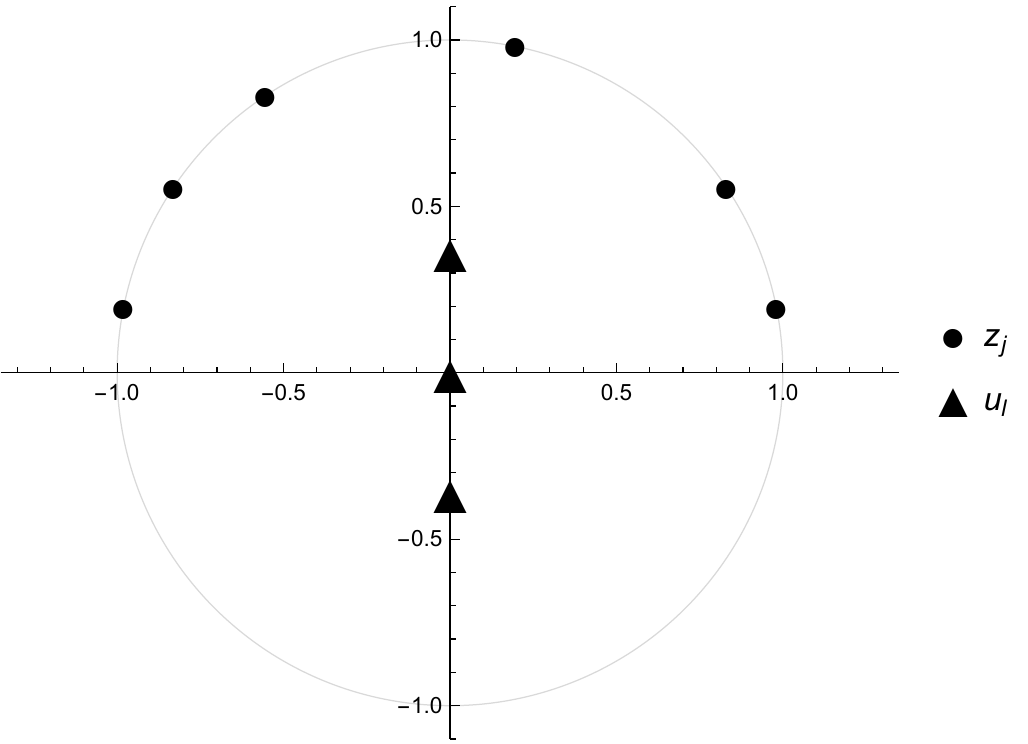}
	\captionsetup{width=0.7\linewidth}
  \captionof{figure}{$L=16$, $(6,\,3)$ sector, $\Lambda=6.613$, free fermionic solution. The six $z_j$'s take six values of the $8^{\text{th}}$ unit roots. Two $u_l$'s form a zero mode Cooper pair, hence they are imaginary and each others negative.}
  \label{fig:BetheRoots1}
\end{minipage}%
\begin{minipage}{.5\textwidth}
  \centering
  \includegraphics[width=.7\linewidth]{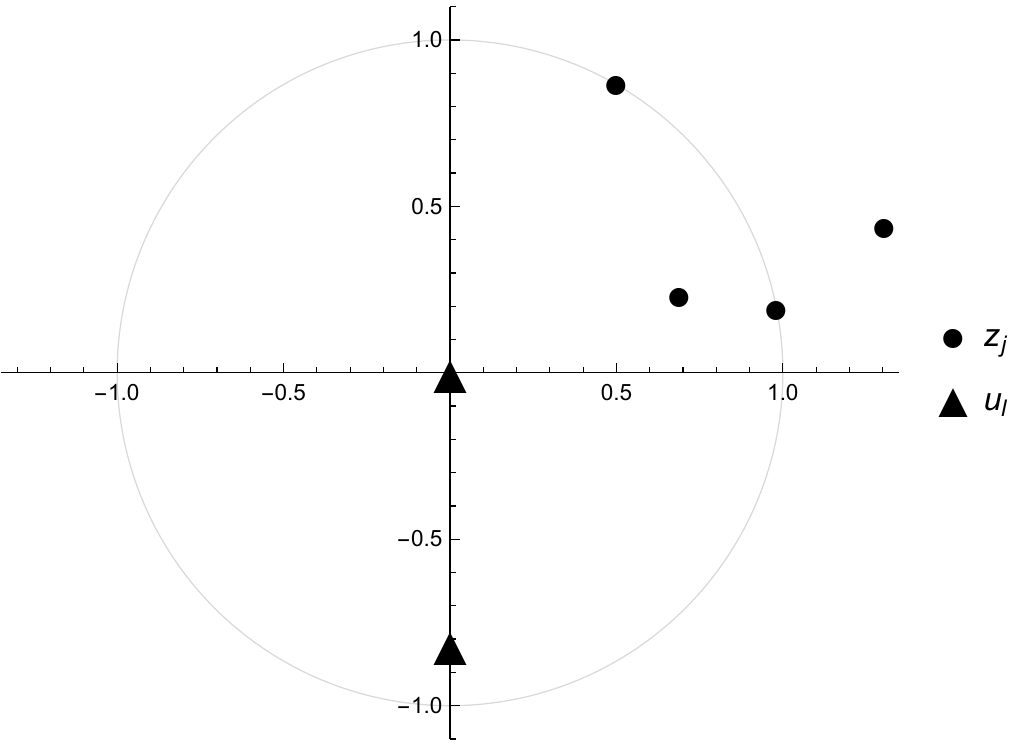}
	\captionsetup{width=0.7\linewidth}
  \captionof{figure}{$L=10$, $(4,\,2)$ sector, $\Lambda=6.721$, non free fermionic solution. The $z_j$'s not on the unit circle are related as $z\to (z^*)^{-1}$. }
  \label{fig:BetheRoots3}
\end{minipage}
\end{figure}

A solution to the Bethe equations gives rise to an eigenvector, however this correspondence is not unique. Two solutions ${z_1, \ldots, z_m, u_1, \ldots, u_k}$ and ${z'_1, \ldots, z'_m, u'_1, \ldots, u'_k}$ give rise to the same eigenvector if there are permutations $\pi \in S_m$, $\sigma \in S_k$  and signs $\epsilon_1, \ldots, \epsilon_m$, $\epsilon_j \in \{+,-\}$ such that $z_j =\epsilon_j z'_{\pi(j)}$ and $u_l = u'_{\sigma(l)}$. In other words, the solutions are invariant under permutations of the Bethe-roots, and invariant under the sign of $z$'s.

Note, that the eigenvalue $\Lambda$ is only dependent on the $z_j$'s. In the absence of odd domain walls, i.e. $k=0$, the equations become free-fermion and are solved by 
\be
z_j = \ii^{-1/2} \e^{\frac{2\ii \pi I_j}{L}},\qquad j=1,\ldots,m
\label{eq:FFsol}
\ee 
where $I_j$ is a (half-)integer. This same solution \eqref{eq:FFsol} can be used to find a solution in the sector with $k=2$ for any solutions $u_1$ and $u_2$ of \eqref{eq:bae2},
\be
1 =  \prod_{j=1}^{m} \frac{u-(z_j-1/z_j)^2}{u+(z_j-1/z_j)^2},
\label{eq:bae_u}
\ee
that are each others negatives, i.e. $u_2=-u_1$. In this case the product in (\ref{eq:bae1}) is
\begin{align}
\frac{u_1-(z_j-1/z_j)^2}{u_1+(z_j-1/z_j)^2}\times\frac{u_2-(z_j-1/z_j)^2}{u_2+(z_j-1/z_j)^2} =\frac{u_1-(z_j-1/z_j)^2}{u_1+(z_j-1/z_j)^2}\times\frac{u_1+(z_j-1/z_j)^2}{u_1-(z_j-1/z_j)^2}=1,
\end{align}
 for any $z_j$, so that \eqref{eq:bae1} with $k=0$, i.e. \eqref{eq:FFsol}, is unchanged.

\begin{figure}[t]
\begin{center}
\includegraphics[scale=0.7]{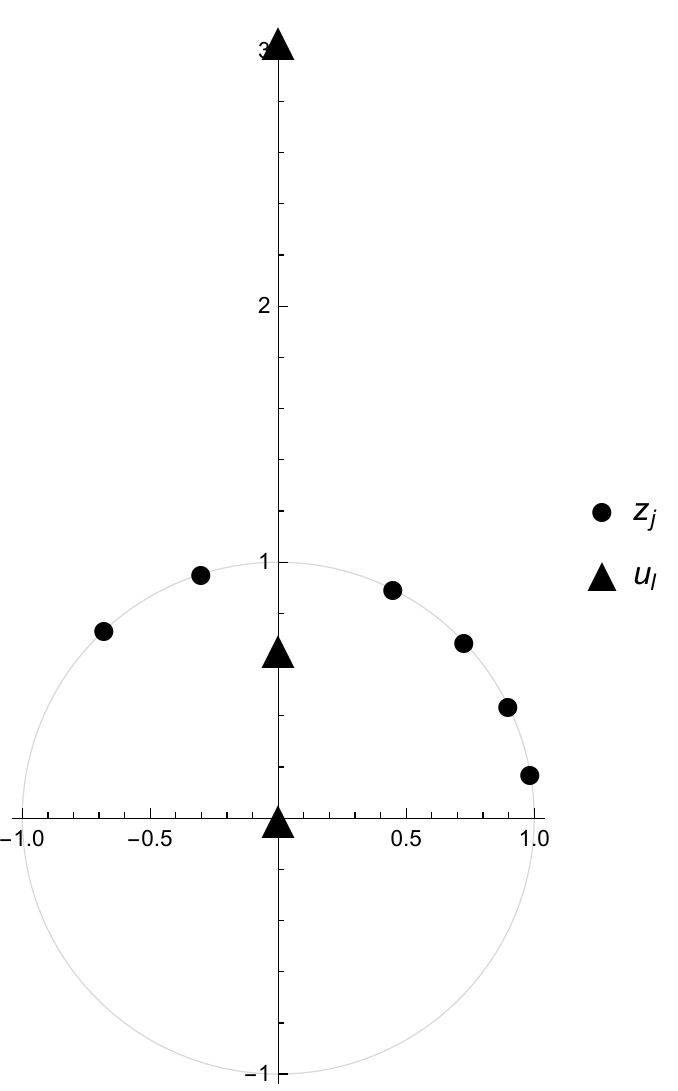}
\caption{$L=16$, $(6,\,3)$ sector, $\Lambda=2.226$, non free fermionic solution on the complex plane. The six $z_j$'s are on the unit circle, but do not take the values of the $8^{\text{th}}$ unit roots.}
\label{fig:BetheRoots2}
\end{center}
\end{figure}

 We can continue like this as long as $m$ is large enough to generate new solutions from \eqref{eq:bae_u}, and add (Cooper) pairs $(u_l,-u_l)$ without changing  the eigenvalue. A similar construction is also possible if we started in a non-free-fermion sector with $k\neq 0$. In sectors where the total number of domain walls $m$ is proportional to the system size $L$ this give rise to an extensive degeneration of energy levels, as we explain in detail in the next section. Some typical solutions to the Bethe equations are shown in Fig.~\ref{fig:BetheRoots1}, \ref{fig:BetheRoots3}, \ref{fig:BetheRoots2}.

We have not been able to find an explicit operator that creates a Cooper pair when acting on a state that admits this.  If such an operator can be constructed, it must either select one of the solution pairs $(u,-u)$ of (\ref{eq:bae_u}), or more likely create a linear combination of all such solution pairs.  Since the pairs do not affect the energy, such linear combination is an eigenstate of the Hamiltonian, but not a pure Bethe state. 

\section{Consequences of symmetry}
\label{sec:Consequences} 
We find that not all the eigenvectors of $H$ are directly described by the Bethe ansatz. However, in all the finite size cases that we looked at, all the eigenvectors were found by applying the symmetry operations on known Bethe vectors. 

The translation symmetry $T$ maps $(m,\,k)$ into $(m,\,m-k)$, eigenvector into eigenvectors. Also, $E$, the DW--nonDW symmetry maps $(m,k)$ into $(L-m,L/2-k)$. By applying both consecutively, $(m,\,k)$ is mapped into $(L-m,\, L/2-m+k)$. The process of the construction of all the eigenvectors from the Bethe vectors is complicated, and we did not find the general structure. Here and in Appendix~\ref{app:GSdeg} we report on certain cases that we studied. \\

\subsection{$L=6$, full spectrum}

For $L=6$, the problem  is easily solvable by direct diagonalisation of the Hamiltonian. According to this, there are four energy levels, all four are $16$-folded degenerate (Table~\ref{tab:L6energies}). 

\begin{table}[h]
\begin{center}
\begin{tabular}{l | l}
$\Lambda$ & deg. \\ \hline
$0.268$ & $16$ \\
$2.000$ & $16$ \\
$3.732$ & $16$ \\
$6.000$ & $16$
\end{tabular}
\caption{$L=6$ sector energy levels and degeneracies.}
\label{tab:L6energies}
\end{center}
\end{table}

In the Bethe ansatz we discriminate even and odd domain walls with an additional nested Bethe root $u_l$ because the interaction between walls depends on the parity of the distance between domain walls. But because it only depends on the distance between domain walls, it makes no difference if we change the parity of all domains walls. In other words, associating a nested Bethe root to the odd DWs is an artifical choice. We therefore identify the sector $(m,k)=(2,0)$ with two even domain walls with that of two odd domains walls $(m,k)=(2,2)$. Hence for $L=6$ there are $6$ different sectors which are listed in Table~\ref{tab:L6sectors}. 

\begin{table}[h]
\begin{center}
\begin{tabular}{l | l | l}
$m$ & $k$ & dim. \\ \hline
$0$ & $0$ & $2$ \\
$2$ & $0$,$2$ & $6$,$6$ \\
$2$ & $1$ & $18$ \\
$4$ & $2$ & $18$ \\
$4$ & $1$,$3$ & $6$,$6$ \\
$6$ & $3$ & $2$
\end{tabular}
\caption{$L=6$ sectors. Certain sectors have the same dimensions and are listed in the same line.}
\label{tab:L6sectors}
\end{center}
\end{table}

Because of the DW-nonDW symmetry, it is enough to probe the lower half of the sectors, i.e. those with $m=0$ and $m=2$. Below we find all eigenvectors corresponding to the dimensions of the eigenspaces given in  Table~\ref{tab:L6sectors}.

\subsubsection{$\Lambda=6$}
The $(m,k)=(0,0)$ sector contains two trivial Bethe vectors: $b_1=\ket{000000}$ and $b_2=\ket{111111}$, both are eigenvectors with $\Lambda = 6$. These vectors are mapped to the (6,3) sector by the DW -- nonDW exchange $E$, to the (2,1) sector by $Q$ and to the (4,2) sector by the combined action of $E$ and $Q$, giving rise to eight vectors: $\{b_1,b_2,Qb_1,Qb_2,Eb_1,Eb_2,Q^\dag Eb_1= EQb_1, Q^\dag E b_2 =EQb_2\}$.

The other eight eigenvectors of this eigenvalue come about in the following way. In the (2,1) sector the Bethe equations are
\begin{align}
z_j^6 & = \pm\ii^{-3} \frac{u-(z_j-1/z_j)^2}{u+(z_j-1/z_j)^2},\qquad j=1,2,\label{eq:baeLambda6a}\\
1 &=  \prod_{j=1}^2 \frac{u-(z_j-1/z_j)^2}{u+(z_j-1/z_j)^2}.\label{eq:baeLambda6b}
\end{align}
Due to Pauli exclusion principle, only distinct pairs $(z_1^2,z_2^2)$ of solutions of \eqref{eq:baeLambda6a} and \eqref{eq:baeLambda6b} give rise to different eigenvectors. In the (2,1) sector there are two independent non-free fermion solutions ($u\neq 0$ and $u\neq\infty$) with $\Lambda=6$, namely 
$$(z_1^2,z_2^2,u_\pm)=(\frac{\sqrt{3}}{2}(\sqrt{3}+\ii), \frac{1}{2\sqrt{3}}(\sqrt{3}+\ii),\frac{1}{39}(-9\pm14\sqrt{3})).$$ 
If we denote the corresponding two Bethe vectors by $b_3$ and $b_4$ then in the (2,1) sector we have the four vectors $\{b_3,b_4,Q^\dag Eb_3=EQb_3,Q^\dag Eb_4=EQb_4\}$ and in the (4,2) sector we find $\{Qb_3,Qb_4,Eb_3,Eb_4\}$. 
 
 In summary we have recovered the full sixteen-dimensional $\Lambda=6$ eigenspace.
 
 \subsubsection{Other eigenvalues}

Based on direct diagonalisation, the $(m,k)=(2,0)$ and $(m,k)=(2,2)$ sectors each contain two eigenvectors associated to each of the lower three eigenvalues. These are reproduced by the Bethe roots in the $(m,k)=(2,0)$ sector, as these satisfy the equation
\begin{equation}
 z_j^6 = \pm \ii^{-3}, \qquad j=1,2.
 \label{eq:baeL6}
\end{equation}
There are precisely two times three distinct pairs $(z_1^2,z_2^2)$ of allowed solutions for the $+$ and $-$ solution respectively, giving each of the lower three eigenvalues twice, and this is doubled using the combined action of $E$ and $Q$. Similarly for $(m,k)=(2,2)$ and by symmetries also in the sectors $(m,k)=(4,1)$ and $(m,k)=(4,3)$. Hence we obtain eight vectors each for the first three eigenvalues. This leaves $24=3\times 8$ vectors still to be determined, and they all must come from the remaining twelve dimensions of the $(m,k)=(2,1)$ (four of the eighteen available vectors in this sector contribute to $\Lambda=6$), as well as the twelve remaining dimensions of the $(m,k)=(4,2)$ sector. 

In the $(m,k)=(2,1)$ sector we may distinguish two types of solutions, the free fermionic (FF) and the non free fermionic (nonFF). The latter we found correspond to $\Lambda=6$, and the FF solutions are those with $u=0$ and $u=\infty$. For $u=0$, we obtain the following BEs,
\begin{align}
z_1^6 = - \ii^{-3},\qquad z_2^6 = - \ii^{-3},
\end{align}
while with $u=\infty$, we find
\begin{align}
z_1^6 = \ii^{-3},\qquad z_2^6 = \ii^{-3},
\end{align}
which are the same as for the $(m,k)=(2,0)$ sector. By the same reasoning as for \eqref{eq:baeL6}, these two sets each produce six solutions, i.e. twelve in total, and by DW-nonDW symmetry we obtain all of the remaining 24 solutions.

We have thus found the complete spectrum for $L=6$ from the Bethe equations and the symmetries.

\subsection{$L=10,\, \Lambda = 6$ degeneracy}

As an other example, we probed the mostly degenerate case in $L=10$, the $\Lambda = 6$ eigenvalue, which is 64-fold degenerate. Because of the DW-nonDW symmetry it is enough to look at the sectors $(m,k)$ with $m<L/2=5$. The Hamiltonian is easily diagonalisable in these sectors giving rise the degeneracies shown in Table~\ref{tab:L10deg}. 

\begin{table}
\begin{center}
\begin{tabular}{l | l | l}
$m$ & $k$ & deg. of $\Lambda = 6$ \\ \hline
$0$ & $0$ & $0$ \\
$2$ & $0$,$2$ & $4$ \\
$2$ & $1$ & $8$ \\ 
$4$ & $0$ & $0$ \\
$4$ & $1$,$3$ & $4$ \\
$4$ & $2$ & $8$ \\
\end{tabular}
\caption{$L=10,\, \Lambda = 6$ degeneracies sector by sector. The unlisted sectors follow by DW-nonDW symmetry.}
\label{tab:L10deg}
\end{center}
\end{table}

The four states in $(2,0)$ are pure Bethe-states and we denote the four-dimensional span of these by $\mathcal{B}^{(2,0)}$. The four states in $(2,2)$ are the copies of these states under the translation symmetry $T$ which shifts all the sites by one. 

Out of the eight states in $(2,1)$, only four are pure Bethe states spanning $\mathcal{B}^{(2,1)}$. Since $Q$ is a symmetry which maps from $(m,\,k)$ to $(m+2,\,k+1)$, by applying $Q$ we create four states each in the $Q\mathcal{B}^{(2,0)}$ subspace of $(4,1)$, the subspace $Q\mathcal{B}^{(2,1)}$ of $(4,2)$, and $QT\mathcal{B}^{(2,0)}$ of $(4,3)$. These all turn out to be linearly independent. 

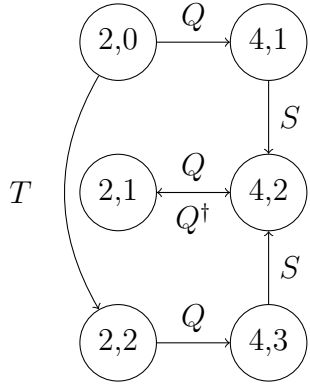
\begin{figure}
\begin{center}
\begin{tikzpicture}
\node[circle,draw](20) at (0,2) {2,0};
\node[circle,draw](41) at (2,2) {4,1};
\draw[->] (20) edge (41);
\node [above] at (1,2){$Q$};

\node[circle,draw](21) at (0,0) {2,1};
\node[circle,draw](42) at (2,0) {4,2};
\draw[->] (21) edge (42);
\node [above] at (1,0){$Q$};
\draw[<-] (21) edge (42);
\node [below] at (1,0){$Q^\dag$};

\node[circle,draw](22) at (0,-2) {2,2};
\node[circle,draw](43) at (2,-2) {4,3};
\draw[->] (22) edge (43);
\node [above] at (1,-2){$Q$};

\draw[->] (43) edge (42);
\node [right] at (2,-1){$S$};
\draw[->] (41) edge (42);
\node [right] at (2,1){$S$};

\draw[->] (20) edge [bend right] (22);
\node[left] at (-1,0) {$T$};
\end{tikzpicture}
\end{center}
\caption{Action of symmetries between domain wall sectors}
\label{fig:L10symmetries}
\end{figure}

$S$ is a symmetry operator which moves one of the the domain walls by one unit, so it maps a state in the sector $(m,\,k)$ into $(m,\,k-1)$ and $(m,\,k+1)$, possibly creating a zero vector. By applying $S$ on $Q\mathcal{B}^{(2,0)}$ we can create two linearly independent (and two linearly dependent) vectors in $(4,2)$, and by applying $S$ on $QT\mathcal{B}^{(2,0)}$ we create the missing two linearly independent vectors (and again two linearly dependent).  Applying $Q^\dagger$ on these four new vectors created by $S$, we found the missing four linearly independent vectors in $(2,1)$. We thus found thirty two states and using the DW-nonDW symmetry we find all sixty-four. This process is depicted in Fig.~\ref{fig:L10symmetries}. 

It would be very interesting to find the full underlying symmetry algebra, i.e. the general algorithm to create all the eigenvectors for given system size and given energy. This may be challenging as it seems not obvious which symmetries create nonzero and linearly independent vectors. 

\subsection{The groundstate for $L=4n$}
\label{sec:L4nGS}

In the half-filled sector $(2n,\,0)$ with $L=4n$ where $2n$ is the total number of domain walls, the Bethe equations
\begin{equation}
 z_j^{4n} =\pm \ii^{-4n/2} = \pm (-1)^n =\pm 1\quad (j=1, \ldots, 2n)
 \label{eq:BeqFFL4n}
\end{equation}
satisfied by the free fermion solutions
\begin{align}
 z^{(+)}_j = \e^{\frac{\ii \pi j}{2 n}} \quad (j=1, \ldots, 2n),\qquad
 z^{(-)}_j = \e^{\frac{\ii \pi (2j+1)}{4 n}} \quad (j=1, \ldots,2n). 
\label{eq:ffsol}
\end{align}
These solutions produce a groundstate as for each of them the eigenvalue
\be
\Lambda=4n+\sum_{j=1}^{2n} (z_j^2+z_j^{-2} -2) = 0.
\ee
These solutions span the sector $(2n,\,0)$, which is also spanned by the two vectors $\ket{0011 \ldots 0011}$ and $\ket{1100 \ldots 1100}$, hence giving these groundstates in terms of Bethe states.

Based on these solutions, we can construct further eigenstates in the sectors $(2n,\,k)$. 
In the presence of $k$ odd DWs, we have 
\begin{equation}
1=\prod_{j=1}^{2n} \frac{u_l-(z_j-1/z_j)^2}{u_l+(z_j-1/z_j)^2}\quad (l=1,\ldots,k).
\label{eq:u}
\end{equation}
After substituting the free fermion solution \eqref{eq:ffsol} into the right hand side of \eqref{eq:u}, the resulting equation for $u$ has purely imaginary roots that form complex conjugate pairs. The key observation is, that the Bethe equations of the $(2n,\, 2k)$ sector can be satisfied with the free fermionic solution (\ref{eq:ffsol}, if we choose the solutions for $u_l$ in (purely imaginary) complex conjugate pairs, as for such a pair we have that $u^*=-u$ so that for each $j$
\be
\frac{u-(z_j-1/z_j)^2}{u+(z_j-1/z_j)^2}\;\frac{u^*-(z_j-1/z_j)^2}{u^*+(z_j-1/z_j)^2}=1.
\ee
Hence the Bethe equations \eqref{eq:bae1} remain of the free fermion form \eqref{eq:BeqFFL4n} for such solutions. This mechanism of zero energy Cooper pairs results in an overall degeneracy for the sector $m=2n$ growing exponentially in $L$. The computation for a lower bound of the growth is in Appendix~\ref{app:GSdeg}.

%
%
%
%

\subsection{The first excited state for $L=4n$}

Based on direct diagonalisation of the Hamiltonian for $L=4,\,8,\,12$, we observe that the first excited states occurs in the sectors $m=2n \pm 2$ with $k$ arbitrary, and $m=2n$ with $k\neq 0,2n$. Since the $(2n-2,\,0)$ sector is purely free fermionic, the Bethe equations are trivial and we can easily determine the first excited state energy for any $L=4n$. This computation is based on the assumption, that the identified free fermion state is indeed the first excited state for any system size. In case it does not hold, the results are an upper bound for the first excited state energy.

The Bethe equations for the $L=4n$, $(2n-2,\,0)$ sector reads,
\begin{equation}
 z_j^L = \pm \ii^{-L/2} = \pm (-1)^n =\pm 1, \quad (j=1, \ldots, 2n-2)
\label{eq:exc}
\end{equation}
These are the same equations as (\ref{eq:BeqFFL4n}), so the independent solution are (\ref{eq:ffsol}). The only difference compared to the groundstate is, that for the groundstate, we had to select all independent Bethe roots, while now we should leave out two, 
\be
\Lambda = 4n + \sum_{i=1}^{2n-2} z_i^{2} + z_i^{-2}. 
\ee

To minimise the energy, we have to minimize $\sum_i z_i^2 + z_i^{-2}$, which is the same as leaving out the two Bethe roots contributing the most. The two largest contributing Bethe roots are $z_{2n}^{(+)}=1$, $z_{1}^{(+)}=\e^{\ii\pi/2n}$ for the $+$ case of \eqref{eq:exc}, and $z_{1}^{(-)}=\e^{\ii\pi /4n}$, $z_{2n-1}^{(-)}=\e^{- \ii\pi/4n}$ for the $-$ case. The two associated energy levels are
\begin{align}
 \Lambda^{(+)}(L=4n) &= 4-2- 2\cos (\pi/n)= 2 (1-\cos (\pi/n)) \\
 \Lambda^{(-)}(L=4n) &= 4-4 \cos (\pi/2n)= 4 (1-\cos (\pi/2n))
\end{align}

As it is easy to see that $\Lambda^{(-)}<\Lambda^{(+)}$ gives the lower energy, hence the first excited state energy. This result correctly reproduces the $L=4,\,8,\,12$ first excited states energies.
The construction above creates the first excited state in the $(2n-2,\, 0)$ sector, but this highly degenerate energy level occurs in many other sectors.

The groundstate energy $\Lambda_{0} = 0$ and therefore the energy gap is given by
\begin{equation}
 \Delta \Lambda_{n} = \Lambda^{(-)}_{n} - \Lambda_{0} = 4 (1-\cos (\pi/2n)) \approx \frac{\pi^2}{2n^2}.
\end{equation}
As we can see, the gap vanishes as $\sim 1/L^2$, which is a sign of a classical diffusive mode. It is worth mentioning that it is a conjecture that for large $n$ the energy $\Lambda_n^{(-)}$ is the first excited level, strictly speaking it is an upper bound.

\section{Bethe ansatz}
\label{sec:BetheAnsatz} 

In this section we give a detailed derivation of the eigenvalues and eigenvectors using the coordinate Bethe ansatz. We have not been able to identify our model with one of the known solvable lattice models that exist in the literature.

As the space of fermions naturally breaks up into sectors labelled by numbers of domain walls, we now introduce a new labelling of the states instead of the fermionic Fock space notation $\ket{\mathbf{\tau}}$. Let $1\le x_1 < x_2 < \ldots, x_m \le L$ denote the positions of $m$ domain walls, then 
\be
\ket{x_1,\ldots,x_m;p_1,\ldots,p_k}_\epsilon,
\ee
denotes the state with $m$ domain walls of which walls $p_1,\ldots,p_k$ are at an odd position. (This notation is convenient for the Bethe ansatz). If the first domain wall is of $01$ type then $\epsilon=0$, otherwise $\epsilon=1$. If all walls are at an even position (or all at and odd position) then the processes \eqref{eq:oddprocess}, involving pairs of domain walls, cannot take place as $x_{i+1} \ge x_i+2$, and the action of the Hamiltonian on a ket state with two domain walls is given by diffusion with hardcore exclusion. 

For clarity and definiteness we give the explicit action of $H$ on the sector with two even domain walls. First we introduce the shift operators $S_i^\pm$ 
\be
S_i^\pm \ket{x_1,\ldots,x_i,\ldots,x_m;p_1,\ldots, p_k} = \ket{x_1,\ldots,x_i\pm 1,\ldots,x_m;p_1,\ldots, p_k}.  
\ee 
Then for $x_2>x_1+2$: 
\begin{align}
H \ket{x_1,x_2}_\epsilon &= \left(L-4 + \sum_{i=1,2}(-1)^{i+\epsilon} \left(S_i^{+2}+S_i^{-2}\right)\right) \ket{x_1,x_2}_\epsilon,
\label{eq:m=2k=0gen}\\
H \ket{x,x+2}_\epsilon &= (L-4 +(-1)^\epsilon(-S_1^{-2} + S_2^{+2})\ \ket{x,x+2}_\epsilon.
\label{eq:m=2k=0exc}
\end{align}

Now consider the case where the first domain wall is odd. Again, if the walls are well separated, i.e. $x_2 > x_1+1$, the action of $H$ is that of diffusion:
\be
H \ket{x_1,x_2;1}_\epsilon = \left(L-4 + \sum_{i=1,2}(-1)^{i+\epsilon} \left(S_i^{+2}+S_i^{-2}\right)\right) \ket{x_1,x_2;1}_\epsilon.
\label{eq:m=2k=1gen}
\ee
The equations are the same if the second wall were on an odd position rather than the first. When two walls are close we no longer have hardcore exclusion, but there is a non-trivial interaction between the walls:
\begin{multline}
H \ket{x,x+1;1}_\epsilon = \big(L-2 +(-1)^\epsilon (- S_1^{-2} + S_2^{+2})\ \ket{x,x+1;1}_\epsilon \\
\mbox{} + (-1)^\epsilon (\ket{x-1,x;2}_\epsilon + \ket{x+1,x+2;2}_\epsilon\big).
\label{eq:m=2k=1exc}
\end{multline}
where in the last two terms the second domain wall has become odd.

In this section we diagonalise the Hamiltonian $H$ given by \eqref{eq:modeldef} using Bethe's ansatz. We will assume that $L$ is even and impose periodic boundary conditions. Since the total number of domain walls, $m$, is conserved, as well as the number of odd domain walls, $k$, the Bethe ansatz can be constructed separately within each $(m,k)$-sector. We therefore write general wave functions in the form
\be
\ket{\Psi(m;k)} = \sum_{\{x_i\}} \sum_{\{p_j\}} \sum_{\epsilon=0,1} \psi_\epsilon(x_1,\ldots,x_m;p_1,\ldots, p_k) \ket{x_1,\ldots,x_m;p_1,\ldots, p_k}_\epsilon,
\ee
and derive the conditions for the coefficients $\psi$ such that $\ket{\Psi(m;k)}$ is an eigenfunction of $H$,
\be
H \ket{\Psi(m;k)} = \Lambda \ket{\Psi(m;k)}.
\ee
We start with the simplest sectors, namely those with two domain walls.

\subsection{Two domain walls $(m=2)$}
\subsubsection{No odd wall $(k=0)$}
Assuming $x_1$ and $x_2$ are even, from \eqref{eq:m=2k=0gen} we find that two walls far apart satisfy
\begin{multline}
\Lambda\psi_\epsilon(x_1,x_2) = (L-4)\psi_\epsilon (x_1,x_2) + (-1)^\epsilon \big( -\psi_\epsilon(x_1+2,x_2) - \psi_\epsilon(x_1-2,x_2)\\ \mbox{} + \psi_\epsilon(x_1,x_2+2)  + \psi_\epsilon(x_1,x_2-2) \big) ,
\end{multline}
while from \eqref{eq:m=2k=0exc} it follows $\psi$ satisfies the condition
\be
-\psi_\epsilon(x-2,x-2) + \psi_\epsilon(x,x) =0. 
\ee
These two equations can be satisfied if we make the ansatz
\be
\psi_\epsilon(x_1,x_2) = c_\epsilon \left(A^{12} (\ii^{1-\epsilon} z_1)^{x_1} (\ii^\epsilon z_2)^{x_2} + A^{21} (\ii^{1-\epsilon} z_2)^{x_1} (\ii^\epsilon z_1)^{x_2}\right) ,\\
\ee
where $z_1$ and $z_2$ are some auxiliary complex numbers to be determined shortly. Using this ansatz we find that the eigenvalue $\Lambda$ and amplitudes $A$ satisfy the conditions
\begin{equation}
\Lambda =L+ \sum_{i=1}^2 (z_i^2+z_i^{-2}-2),\qquad A^{12}+A^{21}=0.
\end{equation}
Imposing the periodic boundary condition on an even lattice of size $L$ gives
\begin{equation}
\psi_\epsilon(x,L+2) = \psi_{1-\epsilon}(2,x),
\end{equation}
which results in
\begin{align}
c_0 A^{12} z_2^L &= c_1A^{21} & c_1A^{12} (\ii z_2)^L  &= c_0A^{21} \\
c_0 A^{21} z_1^L &= c_1A^{12} & c_1A^{21} (\ii z_1)^L  &= c_0A^{12} \nonumber
\end{align}
We thus find that $(c_0/c_1)^2 = \ii^L$ and
\be
z_1^{2L} = z_2^{2L} = \ii^{-L}.
\ee

Taking the square root we have two different set of solutions:

\begin{align}
  \frac{c_0}{c_1} &= \ii^{L/2} & \frac{c_0}{c_1} &= -\ii^{L/2} \\
	z_1^L &= -\frac{c_1}{c_0} = -\ii^{-L/2} & z_1^L &= -\frac{c_1}{c_0} = \ii^{-L/2} \label{EvenWallsBE}\\
	z_2^L &= -\frac{c_1}{c_0} = -\ii^{-L/2} & z_2^L &= -\frac{c_1}{c_0} = \ii^{-L/2} \nonumber
\end{align}

The Pauli exclusion principle implies $z_1 \neq \pm z_2$. Two different solutions $(z_1,\, z_2)$ and $(z'_1,\, z'_2)$ are independent 
(the corresponding Bethe vectors are orthogonal), if their squares are not equal up to interchange. Since for every solution $z$, 
$-z$ is also a solution, it is enough to deal with half of the solutions to (\ref{EvenWallsBE}). This gives $2 \binom{L/2}{2}$ different solutions (where the $2$ is coming from the two different set of solutions). The dimension of the $(2,\,0)$ sector is $2 \binom{L/2}{2}$, so we conclude that the Bethe ansatz gives the full solution in this sector.
%

\subsubsection{One odd wall}
\label{se:d2o1}

We consider now the case that the first wall is at an odd position. Two walls far apart do not interact and satisfy the same equation as if both were on even positions, from \eqref{eq:m=2k=1gen}:
\begin{multline}
\Lambda\psi_\epsilon(x_1,x_2;1) = (L-4)\psi_\epsilon (x_1,x_2;1) + (-1)^\epsilon \big(-\psi_\epsilon(x_1+2,x_2;1) - \psi_\epsilon(x_1-2,x_2;1) \\  \mbox{} + \psi_\epsilon(x_1,x_2+2;1)  + \psi_\epsilon(x_1,x_2-2;1)\big).
\label{eq:eo2d}
\end{multline}
When the walls are distance one apart, the eigenvalue equation changes due to the process described in \eqref{eq:m=2k=1exc}. We find in this case that
\begin{multline}
\Lambda\psi_\epsilon(x,x+1;1) = (L-2)\psi_\epsilon(x,x+1;1) + (-1)^\epsilon \big(-\psi_\epsilon(x-2,x+1;1)  + \psi_\epsilon(x,x+3;1) \\ + \psi_{\epsilon}(x-1,x;2) + \psi_{\epsilon}(x+1,x+2;2) \big).
\end{multline}
And so, setting $x_2=x_1+1$ in \eqref{eq:eo2d}, it follows that the wave function has to satisfy
\begin{multline}
2\psi_\epsilon(x,x+1;1) + (-1)^\epsilon \big( \psi_\epsilon(x+2,x+1;1)  - \psi_\epsilon(x,x-1;1) \\ + \psi_{\epsilon}(x-1,x;2) + \psi_{\epsilon}(x+1,x+2;2) \big)=0.
\label{eq:eo10}
\end{multline}
Likewise, considering the case were the second wall is odd, the condition on the wave function results in 
\begin{multline}
2\psi_\epsilon(x,x+1;2) + (-1)^\epsilon \big(\psi_\epsilon(x+2,x+1;2)  - \psi_\epsilon(x,x-1;2)  \\ + \psi_{\epsilon}(x-1,x;1) + \psi_{\epsilon}(x+1,x+2;1) \big)=0.
\label{eq:eo01}
\end{multline}

To solve equations \eqref{eq:eo2d}, \eqref{eq:eo10} and \eqref{eq:eo01} we make the following ansatz
\be
\psi_\epsilon(x_1,x_2;p) = \sum_{\pi\in S_2} B^{\pi_1\pi_2}_\epsilon(p)  (\ii^{1-\epsilon} z_{\pi_1})^{x_1} (\ii^\epsilon z_{\pi_2})^{x_2},
\ee
where, with a view to later generalizations, we take
\begin{align}
B^{\pi_1\pi_2}_\epsilon(p) &= c_\epsilon (-1)^{\lfloor (p+\epsilon-1)/2\rfloor } A^{\pi_1\pi_2} g(u,z_{\pi_{p}}) \prod_{j=1}^{p-1} f(u,z_{\pi_j}),
\label{eq:Bdef}
\end{align}
and
\begin{equation}
\Lambda = L+ \sum_{i=1}^2 (z_i^2+z_i^{-2}-2),\qquad A^{12}+A^{21}=0.
\end{equation}
With this ansatz, the scattering conditions \eqref{eq:eo10} and \eqref{eq:eo01} become the following equations for the functions $f$ and $g$,
\begin{align}
&\sum_{\pi\in S_2} A^{\pi_1\pi_2} \left[ z_{\pi_2} g(u,z_{\pi_1}) (2- z_{\pi_1}^2-z_{\pi_2}^{-2}) -\ii z_{\pi_1} f_\epsilon (u,z_{\pi_1})g(u,z_{\pi_2}) (z_{\pi_1}^{-2}-z_{\pi_2}^2) \right]=0,\\
&\sum_{\pi\in S_2} A^{\pi_1\pi_2} \left[ z_{\pi_2} f_\epsilon (u,z_{\pi_1}) g(u,z_{\pi_2}) (2- z_{\pi_1}^2-z_{\pi_2}^{-2}) -\ii z_{\pi_1} g(u,z_{\pi_1}) (z_{\pi_1}^{-2}-z_{\pi_2}^2) \right]=0.
\end{align}
It can be easily checked that these equation are solved by the functions
\begin{align}
f(u,z) &= \ii \frac{u-(z-1/z)^2}{u+(z-1/z)^2},\\
g(u,z) &= \frac{z-1/z}{u+(z-1/z)^2},
\end{align}
where $u$ is an additional complex number to be fixed by the boundary conditions.

The periodic boundary condition needs to be implemented carefully as it introduces minus signs,
\be
\psi_\epsilon(x,L+2;1) = \psi_{1-\epsilon} (2,x;2),\qquad \psi_\epsilon(x,L+1;2) = (-1)^{\mathcal{N}_F-1}\psi_{1-\epsilon} (1,x;1),
\ee
and since $\mathcal{N}_F$ is odd in this case, these conditions result in
\be
B^{\pi_1\pi_2}_\epsilon (1) (\ii^\epsilon z_{\pi_2})^L = B^{\pi_2\pi_1}_{1-\epsilon} (2),\qquad
B^{\pi_1\pi_2}_\epsilon (2) (\ii^\epsilon z_{\pi_2})^L = B^{\pi_2\pi_1}_{1-\epsilon} (1).
\label{eq:n=2bc}   
\ee
Combining $\epsilon=0$ and $\epsilon=1$ and using \eqref{eq:Bdef} we find that 
\be
c_0/c_1 = \pm \ii^{L/2+1}.
\ee
Finally we obtain from the two cases in \eqref{eq:n=2bc} that
\begin{align}
z_{\pi_2}^L &= -\frac{c_1}{c_0} \frac{A^{\pi_2\pi_1}}{A^{\pi_1\pi_2}} f(u,z_{\pi_2})
= \frac{c_1}{c_0} \frac{A^{\pi_2\pi_1}}{A^{\pi_1\pi_2}} f(u,z_{\pi_1})^{-1}
\end{align}
resulting in
\begin{align}
z_1^L &= \pm\ii^{-L/2} \frac{u-(z_{1}-1/z_{1})^2}{u+(z_{1}-1/z_{1})^2},\\
z_2^L &= \pm\ii^{-L/2} \frac{u-(z_{2}-1/z_{2})^2}{u+(z_{2}-1/z_{2})^2},
\end{align}
%
%
%
%
with consistency condition
\be
1=-f(u,z_1)f(u,z_2)=\frac{u-(z_{1}-1/z_{1})^2}{u+(z_{1}-1/z_{1})^2} \frac{u-(z_{2}-1/z_{2})^2}{u+(z_{2}-1/z_{2})^2}.
\ee
Note that solutions with $u=0$ give a free fermion spectrum.

\subsection{Arbitrary number of walls}

\subsubsection{No odd wall}
As long as all the walls are far apart ($x_{j+1} - x_{j} > 2 \, \forall j$), the wavefunction amplitude satisfies
\begin{multline}
 \Lambda \psi_{\epsilon} (x_1, \ldots , x_m ) = (L-2m) \psi_{\epsilon} (x_1, \ldots , x_m) +\\+ (-1)^{\epsilon} \sum_{j=1}^m (-1)^j \psi_{\epsilon} (\ldots, x_j -2, \ldots ) + (-1)^j \psi_{\epsilon} (\ldots, x_j + 2, \ldots )
\label{eq:ArbEvenZeroOdd}
\end{multline}
If two walls are distance 2 apart, $x_{i+1} = x_i +2$, then $\psi_{\epsilon} (\ldots, x_i, x_{i+1}-2)$ and $\psi_{\epsilon} (x_i+2, x_{i+1}, \ldots)$ are missing from the sum. 

Taking the $x_{i+1} = x_i + 2$ limit in (\ref{eq:ArbEvenZeroOdd}), we get
\begin{equation}
 0 = (-1)^{\epsilon} (-1)^i \psi_\epsilon (\ldots, x_i +2, x_i +2, \ldots ) + (-1)^\epsilon (-1)^{i+1} \psi_\epsilon (\ldots, x_i, x_i, \ldots)
\end{equation}
These equations are solved by the ansatz
\begin{equation}
 \psi_\epsilon (x_1, \ldots, x_m) = c_\epsilon \sum_{\pi \in S_m} A^{\pi} \prod_{j=1}^{m/2} (\ii^{1-\epsilon} z_{\pi_{2j-1}})^{x_{2j-1}} (\ii^\epsilon z_{2j})^{x_{2j}},
\end{equation}
which is the generalization of the case with two even walls. The solution is also a generalization of that case, namely we find, that
\begin{equation}
 \Lambda = L + \sum_{j=1}^m (z_j^2 + z_j^{-2}-2), \quad A^\pi = \text{sign} (\pi).
\end{equation}
Imposing the periodic boundary condition 
\begin{equation}
 \psi_\epsilon (x_2, \ldots, x_m, x_1 +L)  = \psi_{1-\epsilon} (x_1, \ldots, x_m),
\end{equation}
results in one of the next equations
\begin{equation}
 z_j^L = -\ii^{-L/2}, \quad z_j^L = \ii^{-L/2}.
\end{equation}
\subsubsection{One odd wall}
Let $p$ denote the index of the odd wall, and thus $x_{p}$ denotes its position. In analogy with \eqref{eq:eo10} and \eqref{eq:eo01} we have the following equations for the wave function components in the case $x_{p+1}=x_p+1$,
\begin{multline}
2\psi_\epsilon(\ldots,x_p,x_{p}+1,\ldots;p+1) + (-1)^{\epsilon+p-1} \big[\psi_\epsilon(\ldots,x_p+2,x_p+1,\ldots;p+1)  - \psi_\epsilon(\ldots,x_p,x_p-1,\dots;p+1)\\ + \psi_{\epsilon}(\ldots,x_p-1,x_p,\ldots;p) + \psi_{\epsilon}(\ldots,x_p+1,x_p+2,\ldots;p)\big]=0.
\label{eq:eom01}
\end{multline}
and
\begin{multline}
2\psi_\epsilon(\ldots,x_p,x_{p}+1,\ldots;p) + (-1)^{\epsilon+p-1} \big[\psi_\epsilon(\ldots,x_p+2,x_p+1,\ldots;p)  - \psi_\epsilon(\ldots,x_p,x_p-1,\ldots;p)\\ + \psi_{\epsilon}(\ldots,x_p-1,x_p,\ldots;p+1) + \psi_{\epsilon}(\ldots,x_p+1,x_p+2,\ldots;p+1)\big]=0.
\label{eq:eom10}
\end{multline}
There are additional equations when three walls are close together. In the case where $x_{p+2}=x_{p+1}+1=x_p+2$ with $x_p$ even, the eigenvalue equation leads to the condition
\begin{multline}
4\psi_\epsilon(\ldots,x_p,x_{p}+1,x_p+2,\ldots;p+1) + (-1)^{\epsilon+p-1} \big[\psi_\epsilon(\ldots,x_p+2,x_p+1,x_p+2,\ldots;p+1) \\ + \psi_\epsilon(\ldots,x_p,x_p+1,x_p,\ldots;p+1) - \psi_\epsilon(\ldots,x_p,x_p-1,x_p+2;p+1)\\
- \psi_\epsilon(\ldots,x_p,x_p+3,x_p+2;p+1) + \psi_{\epsilon}(\ldots,x_p-1,x_p,x_p+2,\ldots;p) \\- \psi_{\epsilon}(\ldots,x_p,x_p+2,x_p+3,\ldots;p+2)\big]=0.
\label{eq:eoem}
\end{multline}
In the case where $x_{p+2}=x_{p+1}+1=x_p+3$ with $x_p$ even, the eigenvalue equation leads to the condition
\begin{multline}
2\psi_\epsilon(\ldots,x_p,x_{p}+2,x_p+3,\ldots;p+2) + (-1)^{\epsilon+p-1} \big[\psi_\epsilon(\ldots,x_p,x_p+2,x_p+1,\ldots;p+2) \\ - \psi_\epsilon(\ldots,x_p,x_p+4,x_p+3,\ldots;p+2) +\psi_\epsilon(\ldots,x_p+2,x_p+2,x_p+3,\ldots;p+2)\\
-\psi_\epsilon(\ldots,x_p,x_p,x_p+3,\ldots;p+2) - \psi_\epsilon(\ldots,x_p,x_p+1,x_p+2,\ldots;p+1)\\
+ \psi_\epsilon(\ldots,x_p,x_p+3,x_p+4;p+1) \big]=0,
\label{eq:eem}
\end{multline}
and similar for the case $x_{p+2}=x_{p+1}+2=x_p+3$ with $x_p$ even. These equations are automatically satisfied by the solution from Section~\ref{se:d2o1}. Define therefore the one-domain wall nested wave function by
\be
\phi_{p}^{(\epsilon)} (u;\pi) =  g(z_{\pi_{p}}) (-1)^{\lfloor (p+\epsilon-1)/2\rfloor} \prod_{j=1}^{p-1} f (u,z_{\pi_j}).
\ee
Then the $2n$-domain wall ansatz for the wave function with one odd wall is
\be
\psi_{\epsilon}(x_1,\ldots,x_{2n};p) = c_\epsilon \sum_{\pi\in S_n} A^{\pi_1\ldots\pi_{2n}} \phi_{p}^{(\epsilon)} (u;\pi) \prod_{j=1}^n \left[ (\ii^{1-\epsilon} z_{\pi_{2j-1}})^{x_{2j-1}} (\ii^\epsilon z_{\pi_{2j}})^{x_{2j}} \right],\\
\ee
corresponding to the eigenvalue given by
\be
\Lambda =L+ \sum_{i=1}^{2n} (z_i^2+z_i^{-2}-2),
\label{eq:eigvalm2no1}
\ee

with wavefunction amplitudes
\be
A^{\pi_1 \ldots \pi_{2n}} = \sign(\pi_1 \ldots \pi_{2n}).
\ee

Periodic boundary conditions lead to
\be
\psi_\epsilon(x_1,\ldots,x_{2n-1},L+2;p) = \psi_{1-\epsilon}(2, x_1,\ldots,x_{2n-1};p+1), 
\ee
and
\be
\psi_\epsilon(x_1,\ldots,x_{2n-1},L+1;2n) = (-1)^{\mathcal{N}_F-1}\psi_{1-\epsilon}(1, x_1,\ldots,x_{2n-1};1). 
\ee
Since the parity of $\mathcal{N}_F$ is equal to the parity of the number of odd domain walls, we find the following conditions:
\be
c_\epsilon A^{\pi_1\ldots\pi_{2n}} (\ii^\epsilon z_{\pi_{2n}})^L (-1)^{\lfloor (p+\epsilon-1)/2\rfloor} = c_{1-\epsilon} A^{\pi_{2n}\pi_1\ldots\pi_{2n-1}} (-1)^{\lfloor (p+1-\epsilon)/2\rfloor} f(u,z_{\pi_{2n}}),  
\ee
and
\be
c_\epsilon A^{\pi_1\ldots\pi_{2n}} (\ii^\epsilon z_{\pi_{2n}})^L (-1)^{\lfloor (2n+\epsilon-1)/2\rfloor} \prod_{j=1}^{2n-1} f(u,z_{\pi_{j}}) = c_{1-\epsilon} (-1)^{\lfloor (1-\epsilon)/2\rfloor} A^{\pi_{2n}\pi_1\ldots\pi_{2n-1}}.  
\ee	
Using $(-1)^{\lfloor (p+\epsilon-1)/2\rfloor}=(-1)^{\epsilon-1}(-1)^{\lfloor (p+1-\epsilon)/2\rfloor}$, we obtain again
\be
c_0/c_1=\pm \ii^{L/2+1},
\ee
and the following consistency conditions
\begin{align}
&\prod_{j=1}^{2n} f(u,z_j) = (-1)^{n}\quad \Leftrightarrow \quad\prod_{j=1}^{2n} \frac{u-(z_j-1/z_j)^2}{u+(z_j-1/z_j)^2}=1, \\
& z_j^L = \pm \ii^{-L/2-1} f(u,z_j) = \pm \ii^{-L/2} \frac{u-(z_j-1/z_j)^2}{u+(z_j-1/z_j)^2}\qquad (j=1,\ldots,2n).
\end{align}

Recalling the eigenvalue \eqref{eq:eigvalm2no1}, we note that for all sectors with $2n$ domain walls one of which odd, there exist solutions with $u=0$ giving the free fermion part of the spectrum.

\subsubsection{Two odd walls}

The condition equivalent to \eqref{eq:eoem} when three walls are close together, but now with two at odd positions so that $p_2=p_1+2=p+2$, leads to 
\begin{multline}
4\psi_\epsilon(\ldots,x_p,x_{p}+1,x_p+2,\ldots;p,p+2) + (-1)^{\epsilon+p-1} \big[\psi_\epsilon(\ldots,x_p+2,x_p+1,x_p+2,\ldots;p,p+2) \\ + \psi_\epsilon(\ldots,x_p,x_p+1,x_p,\ldots;p.p+2) - \psi_\epsilon(\ldots,x_p,x_p-1,x_p+2;p,p+2)\\
- \psi_\epsilon(\ldots,x_p,x_p+3,x_p+2;p,p+2) + \psi_{\epsilon}(\ldots,x_p-1,x_p,x_p+2,\ldots;p+1,p+2) \\- \psi_{\epsilon}(\ldots,x_p,x_p+2,x_p+3,\ldots;p,p+1)\big]=0.
\label{eq:oeom}
\end{multline}
The analogue of \eqref{eq:eem} is similar. We find that these are satisfied by the following ansatz for the wave function for $2n$-domain wall of which two are at odd positions:
\begin{multline}
\psi_{\epsilon}(x_1,\ldots,x_{2n};p_1,p_2) = c_\epsilon \sum_{\pi\in S_{2n}} A^{\pi_1\ldots\pi_{2n}} \sum_{\sigma \in S_2} B^{\sigma_1\sigma_2} \\\phi_{p_1}^{(\epsilon)} (u_{\sigma_1};\pi) \phi_{p_2}^{(\epsilon)} (u_{\sigma_2};\pi) \prod_{j=1}^n \left[ (\ii^{1-\epsilon} z_{\pi_{2j-1}})^{x_{2j-1}} (\ii^\epsilon z_{\pi_{2j}})^{x_{2j}} \right],
\end{multline}
Here
\[
A^\pi = \sign(\pi),\quad B^{\sigma}=\sign(\sigma).
\]

Implementing periodic boundary conditions gives rise to 
\be
\psi_\epsilon(x_1,\ldots,x_{2n-1},L+2;p_1,p_2) = \psi_{1-\epsilon}(2, x_1,\ldots,x_{2n-1};p_1+1,p_2+1), 
\ee
and
\be
\psi_\epsilon(x_1,\ldots,x_{2n-1},L+1;p_1,2n) = (-1)^{\mathcal{N}_F-1}\psi_{1-\epsilon}(1, x_1,\ldots,x_{2n-1};1,p_1+1). 
\ee
These give rise to $c_0/c_1=\pm\ii^{L/2+2}$ and the final set of Bethe equations is given by
\begin{align}
z_j^L & = \pm\ii^{-L/2} \prod_{k=1,2}\frac{u_k-(z_j-1/z_j)^2}{u_k+(z_j-1/z_j)^2},\qquad j=1,\ldots,2n\\
1 &=  \prod_{j=1}^{2n} \frac{u_k-(z_j-1/z_j)^2}{u_k+(z_j-1/z_j)^2},\qquad k=1,2.
\end{align}

\subsubsection{Arbitrary number of odd walls}

For the general case we find that the Hamiltonian can be diagonalised by the ansatz
\begin{multline}
\psi_{\epsilon}(x_1,\ldots,x_{2n};p_1,\ldots p_m) = c_\epsilon \sum_{\pi\in S_{2n}} A^{\pi_1\ldots\pi_{2n}} \sum_{\sigma \in S_m} B^{\sigma_1\ldots\sigma_m} \\ \prod_{j=1}^m \phi_{p_j}^{(\epsilon)} (u_{\sigma_j};\pi) \prod_{j=1}^n \left[ (\ii^{1-\epsilon} z_{\pi_{2j-1}})^{x_{2j-1}} (\ii^\epsilon z_{\pi_{2j}})^{x_{2j}} \right],
\end{multline}
where we recall that the wave function for one odd domain wall is given by
\be
\phi_{p}^{(\epsilon)} (u;\pi) =  g(z_{\pi_{p}}) (-1)^{\lfloor (p+\epsilon-1)/2\rfloor} \prod_{j=1}^{p-1} f (u,z_{\pi_j}).
\ee
We find that the eigenvalues of the Hamiltonian are given by
\be
\Lambda =L+ \sum_{i=1}^{2n} (z_i^2+z_i^{-2}-2).
\ee
where the numbers $z_i$ satisfy the following equations
\begin{align}
z_j^L & = \pm\ii^{-L/2} \prod_{k=1}^m \frac{u_k-(z_j-1/z_j)^2}{u_k+(z_j-1/z_j)^2},\qquad j=1,\ldots,2n\\
1 &=  \prod_{j=1}^{2n} \frac{u_k-(z_j-1/z_j)^2}{u_k+(z_j-1/z_j)^2},\qquad k=1,\ldots,m.
\end{align}

\section{Conclusion}
We have introduced a new lattice supersymmetric chain in which fermion number conservation is violated. The model turns out to be integrable and we give a detailed derivation of the equations governing the spectrum using coordinate Bethe ansatz. 

The energy spectrum is highly degenerate, all states with a finite density have an extensive degeneracy. This degeneracy is explained by the identification of several symmetry operators, but most significantly by the possibility at each level to create modes that do not cost any energy. These modes are analoguous to Cooper pairs in BCS theory, and our model contains a direct realisation of these which can be explicitly identified in the Bethe equations. 

The class of finite solutions to the Bethe ansatz does not provide all eigenvectors. We give circumstancial evidence that all eigenvectors are obtained by the application of the symmetry operators on Bethe vectors. We furthermore find that the energy gap to the first excited state scales as $1/L^2$ where $L$ is the system size which is a signature of classical diffusion.

\section*{Acknowledgment}
We are grateful for financial support from the Australian Research Council (ARC), the ARC Centre of Excellence for Mathematical and Statistical Frontiers (ACEMS), and the Koninklijke Nederlandse Academie voor de Wetenschap (KNAW). JdG thanks the hospitality of the Galileo Galilei Institute in Florence where part of this work was finalised. We further warmly thank Lisa Huijse, Jon Links and Kareljan Schoutens for discussions.

\appendix


\section{Groundstate degeneracy for $L=4n$}
\label{app:GSdeg}
In this appendix we discuss the degeneracy of $L=4n$ systems in great depth. Our aim is to give a  lower bound of $\mathcal{O}(2^n)$ on the groundstage degeneracy. These observations are based on counting the zero mode solutions built on the groundstate in sector $(2n,\,0)$, described in Section \ref{sec:L4nGS}. 
A solution in sector $(2n,\, k)$ satisfies the next equations:
\begin{align}
z_j^L & = \pm \ii^{-L/2} \prod_{l=1}^k \frac{u_l-(z_j-1/z_j)^2}{u_l+(z_j-1/z_j)^2},\qquad j=1,\ldots,2n 
\label{eq:GSdeg1}\\
1 &=  \prod_{j=1}^{k} \frac{u_l-(z_j-1/z_j)^2}{u_l+(z_j-1/z_j)^2},\qquad l=1,\ldots,k.
\label{eq:GSdeg2}
\end{align}
Plugging in the groundstate solution (\ref{eq:ffsol}) $z_j^L = \pm 1$ gives rise to  consistency conditions, namely 
\be
\prod_{l=1}^k \frac{u_l-(z_j-1/z_j)^2}{u_l+(z_j-1/z_j)^2} = \pm 1
\label{eq:GSdegSelfCons}
\ee 
in (\ref{eq:GSdeg1}), which can be satisfied by $u=0$, $u=\infty$ and by purely imaginary complex conjugate $u$ pairs, because for these
\be
\frac{u-(z_j-1/z_j)^2}{u+(z_j-1/z_j)^2}\;\frac{u^*-(z_j-1/z_j)^2}{u^*+(z_j-1/z_j)^2}=1.
\ee
 
The possible values of $u_l$'s are fixed by (\ref{eq:GSdeg2}) which is a rational function in variable $u$ after plugging in the $z$'s of the groundstate solution.  Whether the solution of (\ref{eq:GSdeg1}) has to satisfy $z_j^L = 1$ or $z_j^L = -1$  depends on the number of domain walls, and if $u=\infty$ is in the solution:
\begin{itemize}
\item $+1$ equation: \begin{itemize}
											\item even $k$,  all the $u_l$'s form complex conjugate pairs
											\item odd $k$,  $u_1=\infty$, the rest of the $u_l$'s form c.c. pairs
											\end{itemize}
											
\item $-1$ equation: \begin{itemize}
											\item even $k$,  $u_1=0,\, u_2=\infty$, and the rest of the $u_l$'s are c.c. pairs
											\item odd $k$,  $u_1=0$ and the other $u_l$'s are c.c. pairs
											\end{itemize}
\end{itemize}
We have to take into account, that the self consistency condition (\ref{eq:GSdegSelfCons}) has different number of solutions depending on $z_j$'s. If $z_j^L=-\ii^{-L/2}$, it has $2 n$ solutions for even $n$'s, and $2n-1$ for odd ones. Out of these solutions,  $2n-2$ are nonzero c.c. pairs, one is the $u=\infty$ solution, and for even number of solutions, $u=0$ is also a solution. 

The self consistency condition 
induced by $z_j^L=+\ii^{-L/2}$ has $2 n-1$ solutions for even $n$'s, and $2n$ for odd ones. Out of these solutions,  $2n-2$ are nonzero c.c. pairs, one is the $u=\infty$ solution, and for odd number of solutions, $u=0$ is also a solution. 
We have to count the number of $u$ solutions in c.c. pairs, as the degeneracy is from the possible choices among them, while we use $u=0$ and $u=\infty$ to ``tune'' (\ref{eq:GSdeg1}) to $\pm 1$. 

In order to construct new groundstate solutions in the $(2n,\,k)$ sectors, we have to find self consistent solutions: we have to find a set of $u_l$'s, which give the expected $+1$ or $-1$ for (\ref{eq:GSdeg1}), and compute the degeneracy case by case. 
We have to distinguish eight cases: even or odd $n$, even or odd $k$, $+1$ or $-1$ equation and discuss these case by case:
\begin{itemize}
 \item $n$ even, $k$ even, $-1$ equation: In order to get the $-1$ equation with even number of $u_l$'s, $u_1 = 0,\, u_2 = \infty$, and the rest form c.c. pairs. The degeneracy is $\binom{n-1}{(k-2)/2}$.

 \item $n$ even, $k$ odd, $-1$ equation: To get the $-1$ equation, $u_1 = 0$, and the other $k-1$ $u_l$'s form c.c. pairs. The degeneracy is $\binom{n-1}{(k-1)/2}$.

 \item $n$ odd, $k$ even, $-1$ equation: To get the $-1$ equation with even number of $u$'s, we should have $u_1 = 0,\, u_2 = \infty$, but in this sector $u=0$ is not a solution. Consequently, there is no consistent solution.

 \item $n$ odd, $k$ odd, $-1$ equation: To get the $-1$ equation with odd number of $u$'s, we should have $u_1 = 0$, but in this sector $u=0$ is not a solution. Consequently, there is no consistent solution.

 \item $n$ even, $k$ even, $+1$ equation: The $k$ $u$'s have to form $k/2$ c.c. pairs. No $0$ or $\infty$ is involved. The degeneracy is $\binom{n-1}{k/2}$.
 
 \item $n$ even, $k$ odd, $+1$ equation: To satisfy the $+1$ equation, $u_1 = \infty$ and the rest form c.c. pairs. The degeneracy is $\binom{n-1}{(k-1)/2}$.

 \item $n$ odd, $k$ even, $+1$ equation: The $k$ $u$'s have to form $k/2$ c.c. pairs. No $0$ or $\infty$ is involved. The degeneracy is $\binom{n-1}{k/2}$.

 \item $n$ odd, $k$ odd, $+1$ equation: To satisfy the $+1$ equation, $u_1 = \infty$ and the rest form c.c. pairs. The degeneracy is $\binom{n-1}{(k-1)/2}$.
\end{itemize}
Based on this, we can count the states in a certain sector. Instead of counting the explicit results, we would like to point out that summing over $k$ results in a degeneracy proportional to $2^n$. The exact number is not so interesting because this is only a partial degeneracy, with other symmetries, we can construct more states, however the exact number seems to be complicated to find.

\newcommand{\arxiv}[1]{\href{http://arxiv.org/abs/#1}{arXiv/#1}}

\end{document}